\documentclass[nofootinbib, twocolumn, prd]{revtex4-1}
\usepackage{hyperref}
\usepackage[utf8]{inputenc}
\usepackage{amsmath}
\usepackage{amssymb}
\usepackage{courier}
\usepackage{listings}
\usepackage{color}
\usepackage{tabularx,colortbl}
\usepackage{graphicx}
\usepackage{fancyref}
\usepackage{tensor}
\usepackage{appendix}

\begin{document}

\title{Environment Dependent Electron Mass and the H$_0$-Tension}
\author{Rance Solomon}
\email{rancesol (at) buffalo.edu}
\author{Garvita Agarwal}
\author{Dejan Stojkovic}
\affiliation{HEPCOS, Department  of  Physics,  SUNY  at  Buffalo,  Buffalo,  NY  14260-1500, USA}

\begin{abstract}
A difference in the electron mass at recombination from today's measured value has been shown to relieve the H$_0$-tension.
In this article we propose a simple mechanism that would allow for such a mass difference.
Similar in form to the Higgs field, we consider a Yukawa coupling between the symmetron scalar field and the electron.
However, the symmetron's vacuum expectation value depends on its environment's matter density which, in principle, would allow its contribution towards the electron mass to differ between recombination and today.
We detail the coupling and discuss possible observables.
\end{abstract}

\maketitle
\flushbottom

\section{Introduction}\label{sect:intro}
The tension in the measured values of H$_0$ between early and late time observations persists and is now above the 4$\sigma$ discrepancy level.
There are a great number of attempts at solving the tension and bringing the two families of observations into agreement, however, in most cases the tension is only relieved and remains unsolved.
The authors of \cite{DiValentino:2021izs} and  \cite{Schoneberg:2021qvd} review many of the presently leading models and ideas for the resolution of the H$_0$-tension, only a subsection of which we focus on here.
We consider a modified recombination history with higher recombination redshift, $z_*$, allowing for a higher H$_0$ than expected from current CMB estimates (see \cite{Liu:2019awo} for discussion on the degneracy between $z_*$ and H$_0$).

Possibly the most straightforward way to allow for a higher $z_*$ is by considering either a larger fine-structure constant, $\alpha_{EM}$, or a larger electron mass, $m_e$, in the early universe such that recombination occurs at a higher temperature.
Both have been extensively studied (see \cite{Hart:2017ndk,Kaplinghat:1998ry,Avelino:2000ea,Battye:2000ds,Avelino:2001nr,Rocha:2003gc,Martins:2003pe,Scoccola:2009xtv,Menegoni:2012tq}) with similar affects on cosmological parameters.
However, as discussed in \cite{Hart:2019dxi}, it is unlikely that a simple variation in $\alpha_{EM}$ will play a significant role in the H$_0$-tension while changes in $m_e$ remain a viable solution thanks in large part to the difference in dependencies that the Thompson scattering cross-section has on the two parameters, $\sigma_T \approx \alpha_{EM}^2/m_e^2$.
And even though the concept of varying fundamental constants has been around at least since the 1980s \cite{Terazawa:1981ga}, to seriously consider a varying $m_e$ as a possible solution to the H$_0$-tension then a convincing mechanism must be provided that would naturally enable $m_e$ to differ between the early and late universe.
Such is the focus of this article.

We consider the existence of the symmetron field, a scalar-tensor theory with a symmetry breaking potential and a universal coupling to the trace of the stress-energy tensor.
Fifth force interactions are a general characteristic of scalar-tensor theories and as such the symmetron features a screening mechanism in over-dense regions to allow agreement with stringent solar system based measurements of gravity.
We propose a Yukawa interaction between the electron and symmetron such that the screening mechanism has indirect control over the mass of the electron as well.
Much like the Higgs coupling to the electron, the symmetron coupling would naturally give a mass contribution to the electron proportional to the symmetron's vacuum expectation value (VEV)\footnote{Similar principles have been previously discussed in \cite{Olive:2007aj} and more recently in \cite{Fung:2021fcj}.}, $\nu$.
However, unlike the Higgs coupling, the symmetron's VEV depends on the background matter density of its environment.
In regions of high density $\nu=0$ while in low density regions it pulls off towards a finite value.
With the Yukawa coupling, the electron mass would have an additional contribution which varies with the density of its environment.
Therefore, an electron in a high density region (i.e. in the early universe) would have a different mass than one in a low density region (i.e. in low redshift interstellar gas clouds), enabling the necessary electron mass difference between recombination and that observed in galaxy based measurements including those in our solar system.
In general a Yukawa interaction of this sort should be shared by other Standard Model particles.
For simplicity we focus on the electron interaction only.

The prospects of this mechanism is quite intriguing.
It was studied in \cite{Sekiguchi:2020teg} that a varying electron mass by itself can relieve the H$_0$-tension to a good degree but paired with a non-flat cosmology such as $\Omega_k\Lambda$CDM the tension can be almost entirely satisfied.
Even more intriguing though is the pairing of the work here with that in \cite{Desmond:2019ygn} which considers the effect of screened fifth forces on the cosmic distance ladder with the symmetron being one of the studied models.
The authors show that if the Cepheids used for calibrating the distance ladder are screened while those outside the Milky Way are not then H$_0$ would be biased towards higher values.
A full analysis of the effect of a varying $m_e$ on the individual rungs of the distance ladder would have to be discussed in a dedicated article, but, if any effect is seen in Cepheids it will likely push screened Cepheids towards higher emission frequencies than unscreened ones in line with the arguments of section \ref{sect:effects}.
The luminosities of screened Cepheids would then be further underestimated in comparison to those unscreened causing the same propagating effect on the distance ladder as discussed in \cite{Desmond:2019ygn}.
In total, accounting for symmetron fifth forces would decrease the local estimates of H$_0$ while accounting for the electron-symmetron coupling may both increase CMB estimates and decrease local estimates, relieving the tension three-fold.

Section \ref{sect:symmetron} introduces the basics of the symmetron field and some of the current bounds on its parameters while section \ref{sect:coupling} discusses the details of the coupling necessary for shifting H$_0$ towards higher values.
In section \ref{sect:effects} we explore some possible observables of the coupling and we conclude in section \ref{sect:conclusion}.


\section{Symmetron Field}\label{sect:symmetron}
The symmetron field was originally introduced in \cite{Hinterbichler:2010es} and discussed in length in \cite{Hinterbichler:2011ca} as a screening mechanism for long range fifth force interactions from scalar fields in our solar system.
The authors of \cite{Hinterbichler:2010wu} argue for a possible UV completion of the symmetron although this discussion is beyond the scope of this paper.
In its original formulation the symmetron is described by the effective Lagrangian density
\begin{equation}\label{eq:lagrangian}
    \mathcal{L} = \tfrac{1}{2}\partial_\mu\phi\partial^\mu\phi - \tfrac{1}{2}(\tfrac{\Omega}{\Sigma^2} - \mu^2)\phi^2 - \tfrac{1}{4}\lambda\phi^4
\end{equation}
where $\Omega$ is the background fractional energy density, $\Sigma^2 \equiv \tfrac{8\pi GM^2}{3H_0^2}$, and $M$ is treated as a cutoff scale as defined in the original work.
The $\Omega$ term results from a coupling to the trace of the stress-energy tensor and likewise naturally describes only the background matter density (both baryonic and dark matter), but additional arguments can be made to include contributions from radiation as well (see \cite{Dong:2013swa}).
However, since we are not currently interested in the time before recombination, we will not consider the subdominant radiation contributions.

\begin{figure}
    \centering
    \includegraphics[width=\linewidth]{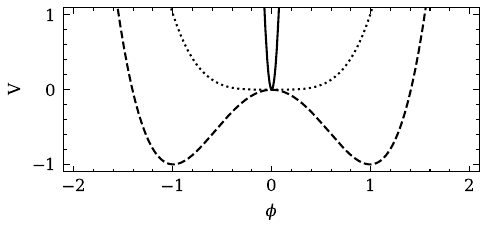}
    \caption{The symmetron potential in arbitrary units. The solid, dotted, and dashed curves show the potential when $\Omega \gg \mu^2\Sigma^2$, $\Omega = \mu^2\Sigma^2$, and $\Omega \ll \mu^2\Sigma^2$, respectively.}
    \label{fig:V}
\end{figure}

In spatial regions where $\Omega \geq \mu^2\Sigma^2$ the mass term of equation \eqref{eq:lagrangian} is positive and $\nu = 0$.
However, when $\Omega < \mu^2\Sigma^2$ symmetry breaking occurs causing the field's mass term to turn negative and the VEV to take a finite value, $\pm \nu \neq 0$ expressed fully in equation \eqref{eq:nu}.
In total,
\begin{equation}\label{eq:nu}
    \nu =
    \begin{cases}
        0 & \Omega \geq \mu^2\Sigma^2, \\
        \sqrt{\tfrac{\mu^2\Sigma^2 - \Omega}{\lambda\Sigma^2}} & \Omega < \mu^2\Sigma^2.
    \end{cases}
\end{equation}
The potential in these two regimes is sketched in figure \ref{fig:V} with arbitrary units.
The transition rate between the solid and dashed curves is dependent on the growth/decay rate of $\Omega$.
In galaxies the transition would be more sensitive to position than time and could be determined using an NFW profile to model $\Omega(r)$ for instance.
\footnote{For simplicity we will not consider the contribution from the primarily baryonic galactic disk (in the case of the Milky Way). Its inclusion would only slightly affect the constraints on $\mu$ and $\Sigma$ due to localized over-densities.}
But throughout the cosmological history the transition would be more sensitive to time or redshift.
This is important to note since we are comparing the electron mass during recombination with the value measured in our solar system which is suspended in a dark matter halo.

\begin{figure}
    \centering
    \includegraphics[width=\linewidth]{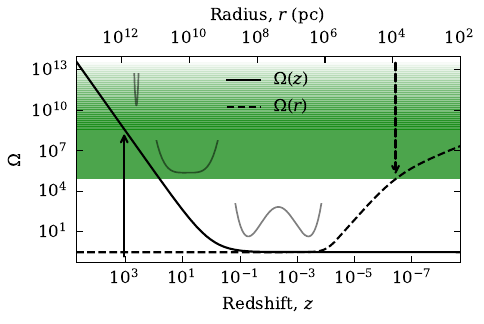}
    \caption{Cosmological matter density versus redshift (solid curve) and the Milky Way matter density versus radius (dashed curve) modeled using an NFW profile with fitting parameters taken from \cite{2019MNRAS.487.5679L}.
    The solid arrow marks recombination while the dashed arrow marks the radial position of the solar system in the Milky Way.
    The green band highlights the expected values of $\mu^2\Sigma^2$ that would allow for a symmetron induced shift in the electron mass in accordance with the H$_0$-tension.
    The behavior of the potential before, at, and after symmetry breaking is shown in insets.}
    \label{fig:densities}
\end{figure}

Reference \cite{Hinterbichler:2010es} has considered symmetry breaking to occur in recent cosmological history such that unscreened long range interactions of gravitational strength could serve as an effective dark energy.
This is not our interest here so the resulting constraints are relaxed.
Instead, we require a difference in the symmetron's VEV between recombination, which had a density of $\Omega_{rec}=\Omega_{m0}(1+z_*)^3\approx4\times10^8$, and today's local value which we take as $\Omega_{MW}\approx8\times10^4$ calculated from an NFW fit to the Milky Way (see figure \ref{fig:densities}).
This translates to $\mu^2\Sigma^2$ being greater than at least one of the above densities.
Since the Milky Way on average has a lower density out of the two we impose
\begin{equation}\label{eq:lower_bound}
    \mu^2\Sigma^2 > \Omega_{MW}.
\end{equation}
Otherwise, if $\Omega_{rec}$ and $\Omega_{MW}$ were both greater than $\mu^2\Sigma^2$ then the symmetry would not be broken in either case and, as we will see in section \ref{sect:coupling}, the symmetron coupling would not effect the predicted value of H$_0$ (though the regime could still be of general cosmological interest).
On the other hand, if $\mu^2\Sigma^2$ is much larger than $\Omega_{rec}$ then the VEV values between recombination and the Milky Way would be indistinguishable and any effect on H$_0$ would be negligible.
That is to say, for the symmetron to cause a significant enough mass difference between recombination and the measured value we would also require
\begin{equation}\label{eq:upper_faded_bound}
    \mu^2\Sigma^2 \not\gg \Omega_{rec}.
\end{equation}

The authors of \cite{Hinterbichler:2010es} had also used local tests of gravity in order to constrain the parameter space.
Since there have not been any anomalous scalar field effects detected in our solar system between the writings of \cite{Hinterbichler:2010es} and this article the resulting constraints are still of interest here.
Particularly, we will take
\begin{equation}
    M \ll M_{Pl}
\end{equation}
resulting in $\mu^2 \gg H_0^2$.
Since the mass of small fluctuations around $\nu$ goes as $m_0 = \sqrt{2}\mu$ the range of fifth force interactions would fall well inside the Hubble radius and may have a role in structure formation.

The parameter $\lambda$ remains to be bounded.


\section{Coupling to the Electron}\label{sect:coupling}
A coupling between the electron, $\psi$, and the symmetron field, $\phi$, can be introduced, in much the same way as with the Higgs field, through an interaction term of the form
\begin{equation}\label{eq:interaction}
    \mathcal{L}_{int} = - g_s\bar{\psi}\psi\phi
\end{equation}
where $g_s$ is an unknown coupling strength between the electron and symmetron.
If we take the mass contributions of the Higgs and symmetron fields to be $\eta_H$ and $\eta_s$, respectively, then at all times the electron mass should be\footnote{It should be noted that there are currently no constraints on $g_H$.
The proposed Future Circular Collider (FCC-ee) may have the ability to probe the low luminosity $H \rightarrow e^+e^-$ process, possibly limiting additional components to the electron mass, as disscussed in \cite{dEnterria:2021xij}.}
\begin{equation}\label{eq:electron_mass}
    m_e = \eta_H + \eta_s.
\end{equation}
The $\eta_H$ term will take its constant canonical form, $\eta_H = \tfrac{g_H\nu_H}{\sqrt{2}}$, while
\begin{equation}\label{eq:eta}
    \eta_s = \dfrac{g_s\nu}{\sqrt{2}} =
    \begin{cases}
        0 & \Omega \geq \mu^2\Sigma^2, \\
        g_s\sqrt{\tfrac{\mu^2\Sigma^2 - \Omega}{2\lambda\Sigma^2}} & \Omega < \mu^2\Sigma^2.
    \end{cases}
\end{equation}
We note that since we have coupled matter to the Jordan frame metric, the fermion mass term becomes $\L_{ferm.} = m_\psi A(\phi)\bar{\psi}\psi$ and \eqref{eq:electron_mass} would have an additional contributions proportional to $m_e\nu^2/M^2$.
But as long as $|g_s| \gg 10^{-84}/\sqrt{\lambda}$ then we can safely take $\eta_s$ as the dominant contribution and treat \eqref{eq:electron_mass} as the electron mass in the symmetron coupled state.
And although we do not apply the constraint here, if for instance one were to apply gravitational strength fifth force interactions as was done in \cite{Hinterbichler:2011ca} then $|g_s| \gg 5\times10^{-23}$.
In addition, including the interaction term \eqref{eq:interaction}, back reaction effects must be taken into account for the shape of $V_{eff}$.
Since the interaction adds a term linear in $\phi$, the VEV will be shifted away from zero during the unbroken symmetry phase.
There will also be a tilt in the symmetry broken phase creating a false vacuum.
Overall these inclusions do not change the essence of our mechanism so we proceed by ignoring the back reactions and push their discussion to appendix \ref{sec:symmetron_backreactions}.

It was argued in \cite{Hart:2019dxi} that the H$_0$-tension can be alleviated to some degree in the $\Lambda$CDM model with an electron mass at recombination given by
\begin{equation}\label{eq:mass_diff}
    m_{e,rec} = \gamma m_{e,MW}
\end{equation}
where $m_{e,MW}$ is the measured electron mass of 0.51MeV.
The fitted parameter $\gamma$ varied depending on the data sets involved but in all cases it fell in the range $1 < \gamma < 1.02$.
Reference \cite{Sekiguchi:2020teg} considered the range, $0.95 < \gamma < 1.05$, in there fitting with an $\Omega_k\Lambda$CDM model, and have likewise found $\gamma>1$ allows for a higher H$_0$ with $1.05 \lesssim \gamma \lesssim 1.1$ giving exceptional agreement with late time measurements.
In either case, we note that $\gamma$ deviates from unity by a small amount ($<10\%$) and to alleviate the H$_0$-tension we impose $\gamma > 1$.

For concreteness we will consider $\Omega_{rec} \geq \mu^2\Sigma^2 > \Omega_{MW}$ which we emphasize by the darker horizontal green band in figure \ref{fig:densities}.
With this choice, during recombination $\nu = 0$ and the electron mass is solely dependent on its Higgs field interaction, but in the immediate vicinity of the solar system the electron would have an additional mass term.
Combining equations \eqref{eq:electron_mass}, \eqref{eq:eta}, and \eqref{eq:mass_diff} we find
\begin{equation}
    \eta_s = g_s\sqrt{\dfrac{\mu^2\Sigma^2-\Omega_{MW}}{\lambda\Sigma^2}} = -\tfrac{\gamma-1}{\gamma}\eta_H.
\end{equation}
With $\gamma > 1$ the coupling constant would have to be negative, $g_s < 0$, meaning that the symmetron coupling reduces the electron mass.


\section{Cosmological and Astrophysical Side Effects}\label{sect:effects}
An environment dependent electron mass should come with a number of additional effects.
For instance, given that the electron mass is lower in less dense regions then the absorption/emission spectrum in hydrogen gas clouds will be shifted towards lower energies when compared to spectrums from higher density regions.
One would then expect a systematic redshift in the spectrum of hydrogen clouds towards the outer edges of galaxies.
The effect would be independent of the rotation of the galaxy and should be symmetric with the density profile of the galaxy.

The most interesting effect however is the progression of the electron mass through cosmological history.
Particularly, as the cosmological matter density drops off, the symmetron contribution to the electron mass would go as $(1+z)^{1.5}$.
This effect would be observed as a deviation of the Lyman-$\alpha$ BAOs from the $\Lambda$CDM prediction.
Such a deviation may have already been observed by the DES collaboration, \cite{DES:2021esc}.
If we further consider low density environments for the Lyman-$\alpha$ emmitting regions then the deviation from $\Lambda$CDM would be expected to track the symmetron's VEV such that the shift in absorption wavelengths would have an additional $(1+z)^{-1.5}$ dependence.


\section{Conclusion}\label{sect:conclusion}
We have discussed a simple mechanism that would enable the electron mass to vary between recombination and today's measured value.
The density dependent VEV of the symmetron field coupled to the electron serves as a control on the electron mass and shifts it to a lower value today assuming a coupling constant $g_s < 0$.
The higher electron mass at recombination allows for recombination to occur at a higher temperature and pushes back $z_*$ allowing H$_0$ to take higher values than currently estimated by \textit{Planck}.

The parameter space of this model is still fairly open which makes confirmation, and more importantly invalidation, difficult.
However, we note the predicted Lyman-$\alpha$ BAO deviations discussed in section \ref{sect:effects} is within our current abilities to measure.
A redshift dependent shift in absorption wavelengths, separate from that due to expansion, would be strong evidence for this coupling.
Given the simplicity of this solution and its application towards the prominent discussions of the H$_0$-tension it would serve well to explore this mechanism in greater detail.


\section*{Acknowledgements}
R.S. would like to thank Jeremy Sakstein for his very helpful comments in the back reaction effects from the Yukawa coupling. D.S. is partially supported by the US National Science Foundation, under Grant No. PHY-2014021.


\appendix

\section{Symmetron in Rewind}

With minor changes to the original derivation of the symmetron field (see \cite{Hinterbichler:2010es}), particularly by considering $-\mu^2 \rightarrow \mu^2$ and
\begin{equation}
    A(\phi) = \big(1 + \dfrac{\phi^2}{2M^2}\big) \rightarrow \big(1 - \dfrac{\phi^2}{2M^2}\big),
\end{equation}
the Lagrangian density can take the alternative form
\begin{equation}\label{eq:lagrangian_star}
    \mathcal{L} = \tfrac{1}{2}\partial_\mu\phi\partial^\mu\phi - \tfrac{1}{2}(\mu^2 - \tfrac{\Omega}{\Sigma^2})\phi^2 - \tfrac{1}{4}\lambda\phi^4.
\end{equation}
The symmetron field would then undergo a symmetry restoration at $\mu^2\Sigma^2$, instead of a symmetry breaking, making the VEV zero at low densities and finite at large densities.
The Yukawa coupling would then allow for an electron mass contribution from the symmetron in the early universe instead of the late universe.
Taking $\Omega_{MW} < \mu^2\Sigma^2 \leq \Omega_{rec}$ and $g_s > 0$ one can reproduce a larger electron mass at recombination than is measured today just like what was argued in the main text of the article.

The advantage to this approach over that found in the main text is one of preference.
Today's measured electron mass would be solely from the well studied Higgs field and deviations from this would only be seen in regions of very large density which is not necessarily an advantage for experimental purposes.
However, besides possibly the effect on theoretical quark stars, this alternative approach would not display any of the astrophysical effects discussed in section \ref{sect:effects} due to the VEV being zero in most cases.

\section{Back reactions from symmetron-electron coupling}
\label{sec:symmetron_backreactions}

The discussion in section \ref{sect:coupling} ignores the effect of the additional Yukawa coupling on $V_{eff}$.
This was done for the sake of simplicity, but now we explore these effects by adding the Yukawa term, \eqref{eq:interaction}, to the effective potential:
\begin{equation}
    V_{eff} = \dfrac{1}{2}\Big(\dfrac{\Omega}{\Sigma^2} - \mu^2\Big)\phi^2 + \dfrac{1}{4}\lambda\phi^4 + g_s\dfrac{M^2\Omega_e}{m_e\Sigma^2}\phi
\end{equation}
where $\Omega_e$ is the fractional electron density which we take to be roughly $\Omega_e \approx 10^{-3} \Omega$.
The VEV can then be calculated from
\begin{equation}
    \nu^3 + \dfrac{(\Omega-\mu^2\Sigma^2)}{\lambda\Sigma^2}\Big(\dfrac{g_sM^2}{10^3m_e} + \nu\Big) = 0.
\end{equation}
In the early universe where $\Omega \gg \mu^2\Sigma^2$ (and therefore $\Omega \gg \lambda\Sigma^2$) the VEV rests at
\begin{equation}\label{eq:early_backreaction}
    \nu \approx -\dfrac{g_sM^2}{10^3m_e}
\end{equation}
and only deviates drastically from this near the transition point, $\Omega \approx \mu^2\Sigma^2$.
On the other hand, when $\Omega \ll \mu^2\Sigma^2$ the VEV is displaced to some finite value (the details of which are not important here) but now with $sign(\nu)=sign(g_s)$.
Therefore, the actual mechanism is messier than what was discussed in section \ref{sect:coupling} but remains the same in principle since there remains a difference in the VEV during recombination and today.

\begin{figure}
    \centering
    \includegraphics[width=\linewidth]{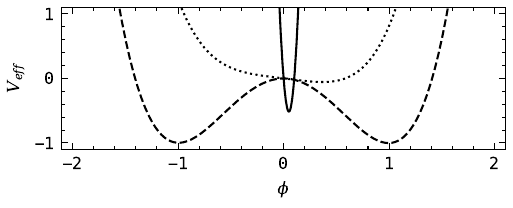}
    \caption{The symmetron potential in arbitrary units including back reaction from the Yukawa coupling (assuming $g_s < 0$). Identical to figure \ref{fig:V}, the solid, dotted, and dashed curves show the potential when $\Omega \gg \mu^2\Sigma^2$, $\Omega = \mu^2\Sigma^2$, and $\Omega \ll \mu^2\Sigma^2$, respectively. For clarity, if tracking $V_{eff}$ through cosmic time one would see a gradual transition from the solid curve, to the dotted, and ending with the dashed.}
    \label{fig:V_br}
\end{figure}

\hspace{0.5cm} With the exception of its affect on parameter estimations, the only physical effects of interest that the back reaction term causes would be in the decreased formation of domain walls.
Without the back reaction term, the field has equal probability of rolling towards $\pm\nu$ assuming the field is initially at rest at $\phi=0$ (see figure \ref{fig:V}).
This would cause a high frequency of domain walls forming between states of $+\nu$ and $-\nu$.
But with the back reaction term, the field at symmetry breaking is already pulled towards one of the VEVs (the true vacuum) while the secondary VEV (false vacuum) is formed at a higher energy (see figure \ref{fig:V_br}).
Domain walls in this case only form when the field $\phi$ tunnels through the potential barrier and are therefore less likely to occur.



\begin{thebibliography}{99}

\bibitem{DiValentino:2021izs}
    E.~Di Valentino, O.~Mena, S.~Pan, L.~Visinelli, W.~Yang, A.~Melchiorri, D.~F.~Mota, A.~G.~Riess and J.~Silk,
    Class. Quant. Grav. \textbf{38}, no.15, 153001 (2021)
    doi:10.1088/1361-6382/ac086d
    [arXiv:2103.01183 [astro-ph.CO]].
    
\bibitem{Schoneberg:2021qvd}
    N.~Sch\"oneberg, G.~Franco Abell\'an, A.~P\'erez S\'anchez, S.~J.~Witte, V.~Poulin and J.~Lesgourgues,
    [arXiv:2107.10291 [astro-ph.CO]].
    
\bibitem{Liu:2019awo}
    M.~Liu, Z.~Huang, X.~Luo, H.~Miao, N.~K.~Singh and L.~Huang,
    Sci. China Phys. Mech. Astron. \textbf{63}, no.9, 290405 (2020)
    doi:10.1007/s11433-019-1509-5
    [arXiv:1912.00190 [astro-ph.CO]].

\bibitem{Hart:2017ndk}
    L.~Hart and J.~Chluba,
    Mon. Not. Roy. Astron. Soc. \textbf{474}, no.2, 1850-1861 (2018)
    doi:10.1093/mnras/stx2783
    [arXiv:1705.03925 [astro-ph.CO]].

\bibitem{Kaplinghat:1998ry}
    M.~Kaplinghat, R.~J.~Scherrer and M.~S.~Turner,
    Phys. Rev. D \textbf{60}, 023516 (1999)
    doi:10.1103/PhysRevD.60.023516
    [arXiv:astro-ph/9810133 [astro-ph]].

\bibitem{Avelino:2000ea}
    P.~P.~Avelino, C.~J.~A.~P.~Martins, G.~Rocha and P.~T.~P.~Viana,
    Phys. Rev. D \textbf{62}, 123508 (2000)
    doi:10.1103/PhysRevD.62.123508
    [arXiv:astro-ph/0008446 [astro-ph]].

\bibitem{Battye:2000ds}
    R.~A.~Battye, R.~Crittenden and J.~Weller,
    Phys. Rev. D \textbf{63}, 043505 (2001)
    doi:10.1103/PhysRevD.63.043505
    [arXiv:astro-ph/0008265 [astro-ph]].

\bibitem{Avelino:2001nr}
    P.~P.~Avelino, S.~Esposito, G.~Mangano, C.~J.~A.~P.~Martins, A.~Melchiorri, G.~Miele, O.~Pisanti, G.~Rocha and P.~T.~P.~Viana,
    Phys. Rev. D \textbf{64}, 103505 (2001)
    doi:10.1103/PhysRevD.64.103505
    [arXiv:astro-ph/0102144 [astro-ph]].

\bibitem{Rocha:2003gc}
    G.~Rocha, R.~Trotta, C.~J.~A.~P.~Martins, A.~Melchiorri, P.~P.~Avelino, R.~Bean and P.~T.~P.~Viana,
    Mon. Not. Roy. Astron. Soc. \textbf{352}, 20 (2004)
    doi:10.1111/j.1365-2966.2004.07832.x
    [arXiv:astro-ph/0309211 [astro-ph]].

\bibitem{Martins:2003pe}
    C.~J.~A.~P.~Martins, A.~Melchiorri, G.~Rocha, R.~Trotta, P.~P.~Avelino and P.~T.~P.~Viana,
    Phys. Lett. B \textbf{585}, 29-34 (2004)
    doi:10.1016/j.physletb.2003.11.080
    [arXiv:astro-ph/0302295 [astro-ph]].

\bibitem{Scoccola:2009xtv}
    C.~G.~Scoccola, S.~J.~Landau and H.~Vucetich,
    Mem. Soc. Ast. It. \textbf{80}, no.4, 814-819 (2009)
    doi:10.1017/S1743921310009476
    [arXiv:0910.1083 [astro-ph.CO]].

\bibitem{Menegoni:2012tq}
    E.~Menegoni, M.~Archidiacono, E.~Calabrese, S.~Galli, C.~J.~A.~P.~Martins and A.~Melchiorri,
    Phys. Rev. D \textbf{85}, 107301 (2012)
    doi:10.1103/PhysRevD.85.107301
    [arXiv:1202.1476 [astro-ph.CO]].

\bibitem{Hart:2019dxi}
    L.~Hart and J.~Chluba,
    Mon. Not. Roy. Astron. Soc. \textbf{493}, no.3, 3255-3263 (2020)
    doi:10.1093/mnras/staa412
    [arXiv:1912.03986 [astro-ph.CO]].

\bibitem{Terazawa:1981ga}
    H.~Terazawa,
    Phys. Lett. B \textbf{101}, 43-47 (1981)
    doi:10.1016/0370-2693(81)90485-8

\bibitem{Hinterbichler:2011ca}
    K.~Hinterbichler, J.~Khoury, A.~Levy and A.~Matas,
    Phys. Rev. D \textbf{84}, 103521 (2011)
    doi:10.1103/PhysRevD.84.103521
    [arXiv:1107.2112 [astro-ph.CO]].
    
\bibitem{Olive:2007aj}
    K.~A.~Olive and M.~Pospelov,
    Phys. Rev. D \textbf{77}, 043524 (2008)
    doi:10.1103/PhysRevD.77.043524
    [arXiv:0709.3825 [hep-ph]].
    
\bibitem{Fung:2021fcj}
    L.~W.~Fung, L.~Li, T.~Liu, H.~N.~Luu, Y.~C.~Qiu and S.~H.~H.~Tye,
    [arXiv:2105.01631 [astro-ph.CO]].

\bibitem{Sekiguchi:2020teg}
    T.~Sekiguchi and T.~Takahashi,
    Phys. Rev. D \textbf{103}, no.8, 083507 (2021)
    doi:10.1103/PhysRevD.103.083507
    [arXiv:2007.03381 [astro-ph.CO]].

\bibitem{Desmond:2019ygn}
    H.~Desmond, B.~Jain and J.~Sakstein,
    Phys. Rev. D \textbf{100}, no.4, 043537 (2019)
    [erratum: Phys. Rev. D \textbf{101}, no.6, 069904 (2020); erratum: Phys. Rev. D \textbf{101}, no.12, 129901 (2020)]
    doi:10.1103/PhysRevD.100.043537
    [arXiv:1907.03778 [astro-ph.CO]].
    
\bibitem{Hinterbichler:2010es}
    K.~Hinterbichler and J.~Khoury,
    Phys. Rev. Lett. \textbf{104}, 231301 (2010)
    doi:10.1103/PhysRevLett.104.231301
    [arXiv:1001.4525 [hep-th]].

\bibitem{Hinterbichler:2010wu}
    K.~Hinterbichler, J.~Khoury and H.~Nastase,
    JHEP \textbf{03}, 061 (2011)
    [erratum: JHEP \textbf{06}, 072 (2011)]
    doi:10.1007/JHEP06(2011)072
    [arXiv:1012.4462 [hep-th]].

\bibitem{Dong:2013swa}
    R.~Dong, W.~H.~Kinney and D.~Stojkovic,
    JCAP \textbf{01}, 021 (2014)
    doi:10.1088/1475-7516/2014/01/021
    [arXiv:1307.4451 [astro-ph.CO]].
    
\bibitem{2019MNRAS.487.5679L}
	H.~Lin, and X.~Li,
	Mon. Not. Roy. Astron. Soc. \textbf{487}, no.4, 5679-5684 (2019)
	doi:10.1093/mnras/stz1698
	[arXiv:1906.08419 [astro-ph.GA]].
	
\bibitem{dEnterria:2021xij}
    D.~d'Enterria, A.~Poldaru and G.~Wojcik,
    [arXiv:2107.02686 [hep-ex]].
    
\bibitem{DES:2021esc}
    T.~M.~C.~Abbott \textit{et al.} [DES],
    [arXiv:2107.04646 [astro-ph.CO]].



\end{thebibliography}
\end{document}